\documentclass[12pt]{article}

\usepackage{graphicx}

\newcommand{\VEV}[1]{\left\langle{#1}\right\rangle}
\newcommand{\lsim} {\buildrel < \over {_\sim}}
\newcommand{\gsim} {\buildrel > \over {_\sim}}
\renewcommand{\bar}[1]{\overline{#1}}
\begin {document}
\begin{flushright}
{\small
SLAC--PUB--11893\\
June 2006\\}
\end{flushright}

\vfill

\begin{center}
{{\bf\large Grand Unification as a Bridge Between String Theory and
Phenomenology}\footnote{This research was supported by the
Department of Energy contract DE--AC02--76SF00515 and by DOE grant
No. DE--FG02--96ER41015.}}

\bigskip \bigskip \bigskip

{Jogesh C. Pati\footnote{E-mail: pati@physics.umd.edu; pati@slac.stanford.edu} \\
Stanford Linear Accelerator Center,
Stanford University, Stanford, CA 94309 \\
Department of Physics, University of Maryland, College Park, MD
20740 }
\medskip
\end{center}

\bigskip

\begin{center}
{\bf\large Abstract }
\end{center}
In the first part of the talk, I explain what empirical evidence
points to the need for having an effective grand unification-like
symmetry possessing the symmetry SU(4)-color in 4D. If one assumes
the premises of a future predictive theory including gravity---be it
string/M theory or a reincarnation---this evidence then suggests
that such a theory should lead to an effective grand
unification-like symmetry as above in 4D, near the string-GUT-scale,
rather than the standard model symmetry. Advantages of an effective
supersymmetric G(224) = SU(2)$_L \times$ SU(2)$_R \times$ SU(4)$^c$
or SO(10) symmetry in 4D in explaining (i) observed neutrino
oscillations, (ii) baryogenesis via leptogenesis, and (iii) certain
fermion mass-relations are noted. And certain distinguishing tests
of a SUSY G(224) or SO(10)-framework involving CP and flavor
violations (as in $\mu \rightarrow e\gamma$, $\tau
\rightarrow\mu\gamma$, edm's of the neutron and the electron) as
well as proton decay are briefly mentioned.

Recalling some of the successes we have had in our understanding of
nature so far, and the current difficulties of string/M theory as
regards the large multiplicity of string vacua, some comments are
made on the traditional goal of understanding {\em vis a vis} the
recently evolved view of landscape and anthropism.

\vfill

\bigskip
\noindent {\it Presented at the Workshop on Einstein's Legacy in The
New Millenium,  Puri, India, December 15--22, 2005  and also at the
symposium ``Under The Spell of Physics" at Gerard 't
Hooft's 60th birthday celebration at Vlieland, Netherlands, July 14--16, 2006. }\\

\vfill \newpage

\setcounter{footnote}{0}

\section{\large Introduction}

The interplay between theory and experiments which probe into the
world of the very small and that of the very large has led to
remarkable progress over the last five decades.  The insights gained
in the two worlds have in fact become linked.  This is indicated
briefly in Chart 1.  Studies of the very small have resolved some
puzzles of the very large, while those of the very large have
constrained our understanding of the very small.

In  this talk, I will mainly discuss the world of the very small
(down to distances $\sim 10^{-16}$ to $10^{-32}$ cm).  In here,
progress in the past half of a century has been especially prominent
in the quest for unification of fundamental particles and their
forces. This had led to the introduction of some unconventional
ideas such as those of electroweak unification~\cite{ref:1a,%
ref:new2,ref:new3}, the color gauge theory along with asymptotic
freedom~\cite{ref:1b} and infrared slavery, quark-lepton unification
together with unification of seemingly different electroweak and
nuclear forces~\cite{ref:1c}-\cite{ref:1f}, baryon-lepton
non-conservation, fermion-boson unification~\cite{ref:1g,ref:1gg},
and finally a grand unity of all matter and all forces including
gravity~\cite{ref:1h,ref:1ha}. These attempts have brought forth in
steps the ideas of the standard model of particle physics, grand
unification, supersymmetry, supergravity, and finally superstring/M
theory with the accompanying notion that space-time has extra
dimensions beyond the familiar four ($d_{\rm total}=10/11$).

\begin{table}
\begin{center}
\includegraphics[width=0.98\textwidth]{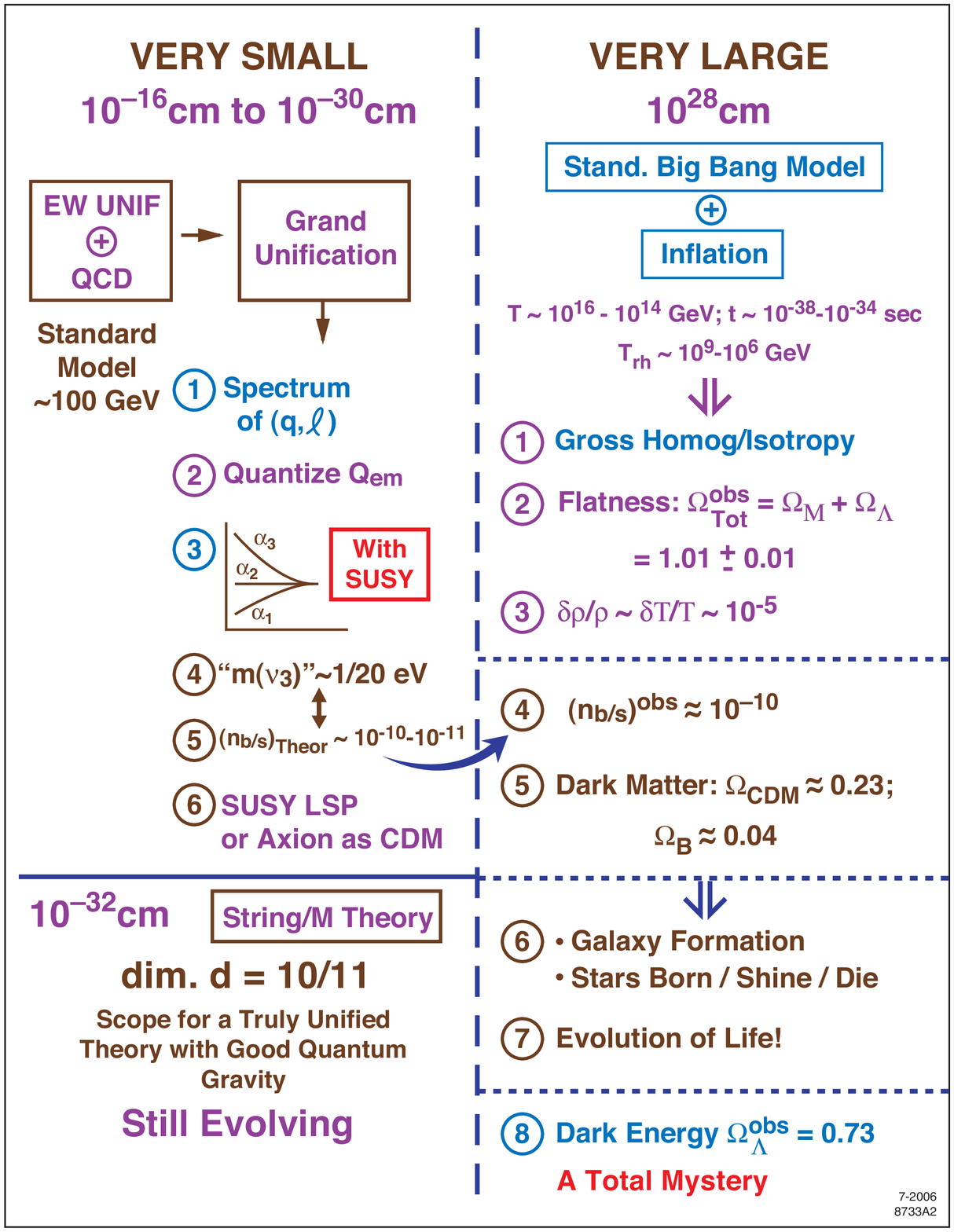}
\end{center}
\noindent {Chart 1:  Some insights gained in the world of the very
small and that of the very large, which have in fact become linked.}
\end{table}

Although I will not discuss much about the world of the very large,
let me at least mention a major idea in this context---that of
inflation~\cite{ref:a}---which evolved in the late 1980s and can go
well with the ideas of supersymmetry and grand unification.  This
idea, presently treated as a paradigm, neatly accounts for some of
the puzzles of cosmology, including (i) the observed large-scale
gross homogeneity and isotropy of the universe, together with small
anisotropies, as well as (ii) almost exact flatness of the universe
($\Omega^{\rm obs}_{\rm TOT}=1.01\, {+0.009 \atop - 0.016}$
\cite{ref:c}, while $\Omega^{\rm Inflation}_{\rm TOT}=1 \pm
\mathcal{O} (10^{-4})$). Inflation in turn requires physics of the
very small, beyond that of the standard model, that can  link very
well with the physics of grand unification. In the process, it also
serves to efficiently dilute, what would have been an embarrassment
for grand unification, that is super heavy topological structures
such as magnetic monopoles.\footnote{This is assuming that inflation
occurred after the monopole production.} I will mention in Section 7
a likely link between inflation and grand unification in connection
with baryogenesis via leptogenesis, leading to $n_b/s \sim
10^{-10}$.

Returning to the various attempts of unification listed above,
string/M theory, contemplating physics at truly short distances
($\sim 10^{-32}$ cm), is the most ambitious of all in that it offers
the scope for unifying all matter and all forces including gravity
and simultaneously providing a good quantum theory of gravity.  But
it is the least understood of all.  As a result, it is yet to make
contact with the real world.\footnote{It does make a non-trivial
prediction, however, that is the existence of gravity.} The
difficulty is that it exhibits, at the present level of exploration,
a very large degeneracy in its vacuum solutions without a guiding
principle to select among them.  This situation may hopefully be
resolved by a better understanding of the theory and/or quite
possibly by the introduction of some radically new ingredients,
which may provide selectivity.

Meanwhile, it is encouraging that there exist string/M theory
solutions in 4D, which, although by no means unique, are at least
semi-realistic in that they possess SM-like or grand
unification-like gauge symmetries with three families.  This
feature, together with the scope for a complete unity offered by the
string/M theory, raises the hope (for many including me) that
string/M theory may well evolve, possibly with the introduction of
new ingredients, so as to describe nature in a predictive manner,
explaining some of its puzzles.  I will briefly comment on these
features at the end.

I will first explain in the next few sections why there is a need
{\em on empirical grounds} to have an effective grand
unification-like symmetry in 4D, that too of a special class, near
the string-scale of $10^{17}-10^{18}$ GeV.  On this basis, I will
argue at the end that an underlying unified theory including
gravity---be it string/M theory or  a reincarnation---as and when it
evolves so as to be predictive---should lead to such a grand
unification-like symmetry in 4D, rather than the SM-symmetry.  That
in turn would serve as a very useful {\em bridge} between string
theory and phenomenology, in accord with observations.

While the contact between string theory and phenomenology is not yet
in sight, fortunately the standard model (SM) of particle physics
and even the ideas of grand unification and low-energy supersymmetry
lend themselves amply to experimental verification; and they also
serve to explain some of the intriguing puzzles of nature.

The SM of particle physics, comprising the notions of electroweak
unification and QCD, is now in excellent agreement with experiments
at least up to energies of order 100 GeV ({\em i.e.} distance scales
$\sim 10^{-16}$ cm).  There are, however, a few observations such as
neutrino oscillations (see discussion in Section 3), baryon
asymmetry, the need for inflation and that for cold dark matter (not
to mention dark energy), which clearly call for new physics beyond
the SM.

The next step in the unification-ladder is the hypothesis of grand
unification~\cite{ref:1c}-\cite{ref:1f}, which proposes an
underlying unity of quarks and leptons and of their three gauge
forces---weak, electromagnetic and strong.  Even though the main
arena of grand unification lies at superhigh energies ($\sim
10^{16}$ GeV) and thus at very short distances ($\sim 10^{-30}$ cm),
where the underlying unity mentioned above is presumed to prevail,
fortunately it does make several intriguing predictions for low
energy physics, some of which have been probed experimentally. These
include:  (i) first and foremost, the quantum numbers of the members
of a family, (ii) quantization of electric charge, (iii) gauge
coupling unification and thereby the weak angle $\sin^2\theta_W$,
(iv) for a certain class of unification symmetries (though not for
all) neutrino oscillations, as observed, and (v) baryon excess via
lepton excess. These probings lend strong support in favor of the
basic idea of grand unification. As we will see, they even serve to
select out the nature of the underlying symmetry. Grand unification
combined with supersymmetry (mentioned below) offers further tests
by which it can be falsified or further vindicated. These include
predictions for processes such as $\mu \rightarrow e\gamma$,
$\tau\rightarrow \mu\gamma$, EDM transitions of the neutron and the
electron, and last but not least, proton decay.  I will mention some
of these predictions towards the end.

Supersymmetry is an idea that evolved in parallel with that of grand
unification in the early 1970s~\cite{ref:1g}.  It proposes a
symmetry between fermions and bosons.  Remarkably enough, it is
needed for consistency of string theory.  And, quite independently,
there is a compelling reason to believe that supersymmetric partners
of quarks and leptons and of the other SM particles should exist
with masses near the TeV-scale, because such a picture seems (in my
opinion) to be the only natural and viable explanation we have so as
to avoid large quantum corrections to the Higgs mass  and thereby
unnatural extreme fine tuning\footnote{Without going into details, I
should mention, however, that low-energy supersymmetry, in general,
poses some generic problems, including: (i) large flavor changing
neutral current (FCNC) transitions; (ii) gravity-induced $d=4$ rapid
proton decay, and (iii) typically large edm's of the neutron and the
electron.  However, there also exist well-motivated ideas on
SUSY-breaking such as gaugino or gauge-mediation (see {\em e.g.} M.
Luty~\cite{ref:i} for a review) which neatly resolve the FCNC
problem (i).  There also exist simple resolutions of (ii) and (iii)
based on symmetries; see {\em e.g.} Pati~\cite{ref:j}, where
symmetries of explicit three family string-derived
solutions~\cite{ref:new16} adequately address (ii). Thus, on
balance, low-energy supersymmetry remains a well-motivated and
viable idea needing direct test at the LHC.}. As a byproduct, such a
SUSY spectrum leads to a dramatic meeting of the three gauge
couplings~\cite{ref:1i,ref:1f}---a feature that in turn supports the
hypotheses of grand unification and low energy SUSY. Such a SUSY
spectrum also provides a natural candidate for cold dark matter that
is needed to account for observations involving the very large.
Fortunately SUSY particles with masses near the TeV-scale can be
searched for at the forthcoming LHC.

Even if the ideas of grand unification and low-energy supersymmetry
are on the right track, it is natural to believe that they arise
from an underlying unified theory that includes gravity and provides
good quantum theory of gravity.  The prime candidate in this regard
is the string/M theory, which is formulated in 10/11 dimensions. The
questions then arise: (1) should such a higher dimensional theory,
upon compactification of the extra dimensions, lead to an effective
theory in 4D near the GUT/string-scale ($\sim 10^{16}-10^{18}$ GeV)
that is SM-like, or grand unification-like?  And (2), if it is the
latter, which of the alternative symmetries---SU(5)~\cite{ref:1e},
or SO(10)~\cite{ref:k}, or an effective symmetry like G(224) =
SU(2)$_L\times$ SU(2)$_R\times $ SU(4)$^c$~\cite{ref:1d}, or
[SU(3)]$^3$~\cite{ref:l}, or flipped SU(5) $\times$
U(1)~\cite{ref:m}---is preferred by data, if any?

I first discuss (in Sections 2--6) why several
empirical observations  strongly suggest the need for new physics
beyond the SM in 4D, of a nature that impressively  matches with the
predictions of grand unification. Furthermore, I will note why
certain observations such as (i) neutrino oscillations, (ii) the
success of leptogenesis as a means for baryogenesis, and (iii) that
of certain fermion mass relations, even serve to select out the
route to higher unification based on the symmetry SU(4)-color in 4D.
This in turn suggests, as the simplest possibility, that the
effective symmetry in 4D near the string/GUT-scale should minimally
be G(214) = SU(2)$_L \times$ I$_{3R} \times$ SU(4)$^c \supset$
G(213), or G(224), or a simple group like SO(10) or
E$_6$~\cite{ref:n}, all of which possess SU(4)-color, as opposed to
alternative symmetries like SU(5) or [SU(3)]$^3$, which do not.

In Sections 2 and 3, I will recall certain desirable features of
symmetries containing SU(4)-color.  In Sections 4--8, I will discuss
briefly some results on CP and flavor violations, and last but not
least proton decay which arise in the same context.

In the concluding part of my talk, covered in Section 10, I will
express a point of view promoting the search for an underlying
unified theory---be it string/M theory or a reincarnation---that
would describe nature in a predictive manner and would explain some
of its presently unexplainable features.  In this regard, I will
provide a perspective inspired by the successes which we have had
over the last 400 years by way of our understanding of nature.  They
include first and foremost the successes of the two theories of
relativity and quantum mechanics.  In the present context, they also
include the successes of the ideas of the standard model, grand
unification and inflation.  Each of these have aided (in varying
degrees) to our understanding of nature.  At the same time,
interestingly enough, each of these have provided certain
ingredients that are crucial to the origin of life.  Based on these
features, I would express a view at the end that leans towards the
traditional approach to understanding as opposed to the recently
evolved view of landscape, combined with anthropism.

\section{\large The Need for Grand Unification with SU(4)-Color in 4D}

The idea of grand unification~\cite{ref:1c}--\cite{ref:1e} was
motivated on aesthetic grounds to achieve a) a unification of quarks
and leptons, and b) a unity of the electroweak and strong forces.
Simultaneously it was inspired by the desire to explain c) the
observed quantum numbers of the members of a family, and d)
quantization of electric charge. Over the years, the evidence
building up in favor of grand unification (including both old and
new) has become strong~\cite{ref:new21}.  It includes:

\begin{enumerate}
\item First and foremost, the spectrum of quarks and leptons in a
family;

\item Quantization of electric charge;

\item $Q_{e^-}/Q_p = -1$;

\item The dramatic meeting of the three gauge couplings;

\item Equally dramatic, neutrino oscillations~\cite{ref:o}-\cite{ref:q} with $\sqrt{\Delta
m^2(\nu)_{23}}\approx 1/20 \ eV$;

\item The success of the two fermion mass relations:
 \begin{eqnarray}
(a)\  m_b(M_{\rm GUT}) \approx m_\tau, {\rm
and}~~~~~~~~~~~~~~~~~~~~~~~~~~~~~~~~~~~~~~~~~~~~~
~~~~~~~~~~~~~~~~~~~~~~~~~~~~~~~~~~~~~~~~~~~~~~~~~~~~~~~~~~~~~~~\nonumber\\
(b)\  m(\nu^\tau_{\rm Dirac}) \approx m_{\rm top} (M_{\rm GUT}) {\rm (needed\ for\ seesaw);\ and}
~~~~~~~~~~~~~~~~~~~~~~~~~~~~~~ (1)
~~~~~~~~~~~~~~~~~~~~~~~~~~~~~~~~~~~~~~\nonumber\label{eq:1} \end{eqnarray}
\setcounter{equation}{1}
\item Successful baryogenesis via leptogenesis leading to~\cite{ref:r,ref:s}
$Y_B \sim 10^{-10}$.
\end{enumerate}
\noindent Of these, the first three are old.  They served as the
driving motivations for grand unification.  The last four pieces of
evidence have emerged subsequently.

As mentioned before, the meeting of the three gauge couplings, based
on accurate LEP data, occurs if one assumes low-energy
supersymmetry~\cite{ref:1i}.  Such a meeting  thus provides a strong
support both for the ideas of grand unification being operative in
4D (or for an underlying theory like string theory ensuring coupling
unification irrespective of a grand unification symmetry in 4D, see
below) and low-energy supersymmetry. The meeting of the gauge
couplings extrapolated from low momenta occurs at a scale $M_{\rm
GUT}\approx 2\times 10^{16}$ GeV.  As we will see, this  scale
$M_{\rm GUT}$ plays a crucial role in our understanding of the tiny
neutrino masses as well as phenomena such as leptogenesis in the
context of inflation.

While the first four features (1)--(4) provide strong support, on
empirical grounds, in favor of grand unification, they leave open
the question of the choice of the effective symmetry $G$ in 4D near
the GUT-scale.  In particular, they do not make a sharp distinction
between the alternatives of (i) SU(5), (ii) SO(10), (iii) $E_6$,
(iv) [SU(3)]$^3$, or (v) a string-derived semi-simple group like
G(224), with coupling unification being ensured in this case by
string theory at the string scale (see remarks below), or (vi)
flipped SU(5) $\times$ U(1). Of these, the symmetries G(224), SO(10)
and $E_6$ possess the symmetry SU(4)-color, while SU(5), [SU(3)]$^3$
and flipped SU(5) $\times$ U(1) do not.

I would argue in this section and the next that the last three
features, involving: (5) Neutrino oscillations, (6) The success of
the two mass relations 1(a) and 1(b), and (7) The success of
baryogenesis via leptogenesis, clearly suggest that the effective
symmetry $G$ in 4D should possess the symmetry SU(4)-color.  I will
mention in a moment the common advantages shared by SO(10) and a
string-derived G(224)-symmetry as well as the distinctions between
them.

\subsection{\normalsize The Family Multiplet Structure of G(224)}

To see the need for having SU(4)-color as a component of the higher
gauge symmetry, it is useful to recall the family-multiplet
structure of G(224), which is retained by SO(10) as well. The
symmetry G(224) = SU(2)$_L \times$ SU(2)$_R \times$ SU(4)$^c$,
subject to left-right discrete symmetry which is natural to G(224),
organizes members of a family into a {\em single} left-right
self-conjugate multiplet ($\rm{F_L^e}\bigoplus\rm{F_R^e}$) given
by~\cite{ref:1d}:
\begin{eqnarray}
\label{eq:fermion}
\begin{array}{c}
{\rm F_{L,R}^e}=\left[
\begin{array}{cccc}
{\rm u_r}&{\rm u_y}&{\rm u_b}&\mathbf{\nu_e}\\{\rm d_r}&{\rm
d_y}&{\rm d_b}&{\rm e^-}
\end{array}\right]_{\rm L,R}
\end{array} \label{eq:2}
\end{eqnarray}
The multiplets $\rm{F_L^e}$ and $\rm{F_R^e}$ are left-right
conjugates of each other transforming respectively as {\bf (2, 1,
4)} and {\bf (1, 2, 4)} of G(224); likewise for the muon and the tau
families. Note that {\it each family of G(224), subject to
left-right symmetry, must contain sixteen two-component objects} as
opposed to fifteen for SU(5)   or the standard model. While the
symmetries $SU(2)_{L,R}\subset G(224)$ treat each column of
$\rm{F_{L,R}^e}$ as doublets, the symmetry SU(4)-color unifies
quarks and leptons by treating each row of $\rm{F_L^e}$ and
$\rm{F_R^e}$ as a quartet. {\em Thus both SU(4)-color and SU(2)$_R$
predict the existence of the right-handed neutrino as an essential
member of each family, with non-trivial SU(4)$^c$ and SU(2)$_R$
quantum numbers~\cite{ref:1d}.} In particular, SU(4)-color treats
the left and right-handed neutrinos ($\nu_L^e$ and $\nu_R^e$) as the
{\em fourth color-partners} of the left and right-handed up quarks
(u$_{\rm L}$ and u$_{\rm R}$) respectively; likewise for the $\mu$
and the $\tau$ families.  This is why SU(4)-color  leads to some
very desirable fermion mass relations for the third family ({\em
i.e.} both Eqs. 1(a) and 1(b)) that are empirically favored.

{\em An accompanying characteristic of SU(4)-color is that it also
introduces B--L as a local symmetry}.  Now, using observed gauge
coupling unification, one can argue that B--L and thus
SU(4)-color---be it part of SO(10) or a string-derived G(224)
symmetry---should break spontaneously near the GUT-scale rather than
near the Planck or an intermediate scale.  Thus,  $M_{\rm B-L}\sim
M_{\rm GUT}$, where $M_{\rm B-L}$ denotes the (B--L)-breaking scale.
That in turn says that the Majorana mass of the RH neutrino (at
least the heaviest one) which necessarily breaks B--L, should be
near the GUT-scale (see Section 3) rather than being much heavier
($\sim M_{\rm Planck}$) or much lighter than the GUT-scale ($\sim$ 1
to 10 TeV, say).  As we will see in Section 3, such a constraint on
the Majorana mass of the RH neutrino, which would not exist if B--L
were not a part of the gauge symmetry (as in SU(5)), plays a crucial
role in yielding the desired mass of the light LH neutrino as well
as in providing successful baryogenesis via leptogenesis.

\subsection{\normalsize Advantages of G(224)}

The symmetry G(224), supplemented by L--R discrete symmetry which is
natural to G(224), brings a host of desirable features.  Including
some of those mentioned above which served as motivations for grand
unification, they are:

\begin{description}
\item{(i)\ } Unification of all sixteen members of a family within {\em
one left-right self-conjugate multiplet} with a neat explanation of
their quantum numbers;
\item{(ii)}
Quantization of electric charge;
\item{(iii)}
$Q_{e^-}/Q_p = -1$;
\item{(iv)}
Quark-lepton unification through SU(4)-color;
\item{(v)}
Conservation of parity at a fundamental
level~\cite{ref:t};\footnote{It appears aesthetically attractive to
assume that symmetries like Parity (P), Charge Conjugation (C), CP
and Time Reversal (T) break only spontaneously like the gauge
symmetries.  While such a preference a priori is clearly subjective,
and is not respected by symmetries such as SU(5), observations of
neutrino oscillations and the likely need for leptogenesis,
suggesting the existence of $\nu_R$'s a la SU(4)-color and SU(2)$_L
\times$ SU(2)$_R$, seem to go well with the notion of exact
conservation of parity at a fundamental level.}
\item{(vi)}
RH neutrino as a compelling member of each family that is now needed
for seesaw and leptogenesis;
\item{(vii)}
B--L as a local symmetry. It has been realized eventually that this
is needed to protect $\nu_R$'s from acquiring Planck scale masses
and to set (for reasons noted above) $M(\nu^i_R) \propto M_{B-L}\sim
M_{\rm GUT}$, both crucial to seesaw and leptogenesis;
\item{(viii)}
The rationale for the two successful mass-relations 1(a) and 1(b),
the first ($m_b (M_{\rm GUT}) \approx m_\tau$) being empirically
successful and the second ($m(\nu^\tau_{\rm Dirac}) \approx m_{\rm
top}(M_{\rm GUT})$) being a crucial ingredient for the success of
the seesaw.
\end{description}

These eight features constitute the {\em hallmark} of G(224).
Historically, all the ingredients underlying these features, and
explicitly (i)--(vii), including the RH $\nu$'s, B--L and
SU(4)-color, were introduced into the literature only through the
symmetry G(224)~\cite{ref:1d}; this was well before SO(10) or (even)
SU(5) appeared. Any simple or semi-simple group that contains G(224)
would of course naturally possess these features.  So does therefore
SO(10), which is the smallest simple group containing G(224). In
fact, {\em all the advantages of SO(10), which distinguish it from
SU(5) and are now needed to understand neutrino oscillations as well
as baryogenesis via leptogenesis, arise entirely through the
symmetry G(224)}.  SO(10) being the smallest extension preserves
even the family-multiplet structure of G(224) without needing
additional fermions (unlike $E_6$).  The L--R conjugate {\bf
16-plet} $= (F_L \oplus F_R)$ of G(224) precisely corresponds to the
spinorial {\bf 16} $=(F_l\oplus (F_R)^c)$ of SO(10).  Furthermore,
with SU(4)-color being vectorial, G(224) is anomaly-free; SO(10) is
anomaly-free as a group.

As we will see, the last three features (vi), (vii) and (viii) are
strongly favored by observations; and they in turn clearly favor the
class of symmetries possessing SU(4)-color (like G(214), G(224),
SO(10) and $E_6$) over those which do not. For instance SU(5),
devoid of SU(4)-color, does not provide the ingredients of (vi)  RH
neutrino, (vii)  B--L, and (viii)  the mass relation 1(b), though it
does provide 1(a). As a result it does not have a natural setting
for understanding neutrino masses and implementing baryogenesis via
leptogenesis (see discussion in Section 3 and especially footnote
g).

\subsection{\normalsize Similarities and Distinctions Between SO(10) Versus a
String-Derived G(224) in 4D}

In the context of an underlying unified theory like string/M theory
defined in higher dimensions ($d= 10/11$) the question arises:
should the four-dimensional symmetry obtained through
compactification of the extra dimensions contain SO(10) or just
G(224)?  Quite clearly both share the advantages (i)--(viii)
primarily because both contain SU(4)-color and quantize electric
charge.

Distinctions between them arise, however, as regards the issues of
(i) gauge coupling unification, (ii) the so-called doublet-triplet
splitting problem, and (iii) proton decay.  I have discussed these
issues in more detail elsewhere~\cite{ref:u}.

Briefly speaking, for the case of a string-derived
G(224)-solution~\cite{ref:v}, coupling unification
$(g_{2L}=g_{2R}=g_4)$ can hold at the string scale
$M_{st}$~\cite{ref:w} through the constraints of string theory even
though G(224) is semi-simple. As mentioned before, one would then
need to assume that the string-scale is not far above the
conventional GUT-scale $(M_{\rm st} \approx$ (2--3) $M_{\rm GUT}$,
say, with $M_{\rm GUT} \approx 2\times 10^{16}$ GeV) where G(224)
should break to G(213) to explain observed gauge coupling
unification~\cite{ref:x}. While such a possibility can well arise in
the string-context~\cite{ref:y}, an SO(10)-solution would have an
advantage in this regard in that it would ensure gauge coupling
unification at the GUT-scale regardless of the gap between the
string and the GUT-scales.

At the same time, as noted by several authors (see {\em e.g.}
Ref.~\cite{ref:new16} and \cite{ref:v}), a G(224)-solution in 4D
would, however, have a distinct advantage over an SO(10)-solution as
regards the problem of doublet-triplet splitting. This is because,
for a G(224) solution, the undesired color-triplets, which could
induce rapid proton decay, can be naturally projected out through
the process of string compactification.  On the other hand, for an
SO(10)-solution the entire 10-plet must exist in 4D.  One must then
invoke a suitable doublet-triplet splitting mechanism in 4D which
would make the color-triplets  in 10$_H$ superheavy, while keeping
the SU(2)-doublets  in 10$_H$ light. Such a mechanism can be
constructed in 4D~\cite{ref:z}; but it is not clear whether such a
mechanism can in fact emerge consistently from a string theory.

Given the relative advantages of SO(10) and a string-derived G(224)
over each other as four-dimensional symmetries, and the fact that
the possible disadvantage in each case has at least a plausible
solution, I have expressed elsewhere~\cite{ref:u} that it is prudent
to keep an open mind at present towards both, especially because
they share the advantages (i)--(viii) and lead essentially to the
same predictions for fermion masses, neutrino oscillations, and
leptogenesis (see Sections 4 and 6).  I will therefore use them
interchangeably for most purposes. I will mention briefly in
Sections 5, 6  and 8 that the two cases can be distinguished
empirically by phenomena involving CP and flavor violations as well
as proton decay.

\section{\large The SuperK Scale, Seesaw and SU(4)-color}

Atmospheric neutrino oscillation discovered at SuperK~\cite{ref:o}
yields a mass-scale \break $\sqrt{\Delta m^2(\nu)_{23}} \approx
1/20$ eV, with an oscillation angle $\sin^2 2\theta^\nu_{23} \approx
0.92-1.$ By using WMAP and other astronomical data, one also knows
that $\sum m(\nu_i) < 0.68$ eV~\cite{ref:c}.  In other words, even
the heaviest among the three light neutrinos is lighter than 0.68
eV.  These non-vanishing but tiny neutrino masses seem to be
extraordinarily small compared with the masses of the charged
fermions.  For example, comparing ratios of masses of fermions in
the {\em same family}, which I think is more appropriate for the
purpose,\footnote{Some may wish to compare the lightest in the first
family with the heaviest in the third family.  Even here, the ratio
$m_{\nu_e}/m_{\rm top} \lsim$ (1/10 eV) /(120 GeV) $\sim 10^{-12}$
is much smaller than $m_e/m_{\rm top} \sim 10^{-5}$.} we see that
the mass-hierarchies among the charged fermions are of order 1/10 --
1/100 ({\em e.g.} $m_b/m_t \sim $ 1/45; $m_\tau/m_t \sim$ 1/70;
$m_s/m_c \sim$ 1/8, $m_\mu/m_c \sim$ 1/3). By contrast, comparing
masses of neutrinos with the heaviest charged fermion in the same
family, we see that the ratios
\begin{equation}
\left[m(``{\nu_\tau}")/m_{\rm top}\right]_{\rm obs} \sim (1/10 -
1/2)\ {\rm eV/120\ GeV} \sim 10^{-12}- 10^{-11}\ , \label{eq:3aa}
\end{equation}
and likewise $m(``{\nu_\mu}")/m_\mu \lsim 10^{-10}-10^{-9}$ are {\em
way too small} compared with the corresponding ratios of the masses
of the charged fermions in the same family, as mentioned above.

While we do not really understand at present the latter {\em i.e.}
intra- and inter-family mass-hierarchies among the charged fermions,
{\em the extraordinarily small neutrino masses seem to suggest that
some characteristically new physics is at play in ensuring the tiny
neutrino masses}.\footnote{Note if neutrinos were strictly massless,
that feature could be understood simply by assuming the
two-component theory of the neutrino (as in the SM or SU(5)) and
lepton number conservation (despite quantum gravity).  Such
masslessness  of the neutrinos, with this reasoning, is in fact what
a large number of physicists believed until the 1980s and even late
1990s. Alternatively, if neutrinos had masses $\lsim 10^{-5}$ eV
(assuming that could be detected), that too can be understood in the
SM and SU(5) by allowing for quantum gravity-induced lepton number
violation (see text).  {\it It is the superK scale $\sim$ 1/20 eV,
that is neither zero nor as small as $10^{-5}$ eV, that cannot be
accounted for naturally by either of these two alternatives.}  It
thus calls for {\em new physics} beyond the SM.} In short, ratios
such as $m(``{\nu_\tau}")/m_{\rm top} \sim 10^{-12}$ pose a major
puzzle. As we will see, understanding this ratio and thus the superK
mass-scale would shed much light on the nature of the underlying
symmetry.

First of all, one can argue that the superK scale clearly calls for
physics beyond those of the SM, even with allowance for quantum
gravity. This is because, with only LH neutrinos in the SM, even if
one allows for lepton number violation through quantum
gravity-effects and thereby permits~\cite{ref:3aa} an effective
interaction of the form $\kappa LL \phi_H \phi_H/M_{pl}$, it would
yield a Majorana mass-term for $\nu_L$ given by $\kappa \,
\nu^T_L\nu_L \VEV{\phi_H}^2/M_{pl}$, and thus a Majorana mass
$m(\nu_L) = \kappa \VEV{\phi_H}^2 /M_{pl} \approx \kappa(250\ {\rm
GeV})^2/2\times 10^{18}\ {\rm GeV} \approx \kappa (3\times 10^{-5}\
{\rm eV})$.  This is too small by a factor of $10^3$ compared to the
SuperK value (one of course does not want $\kappa$ to be much
greater than one). The situation, however, changes dramatically, as
discussed below, when there is a RH neutrino (in addition to the LH
one) in the context of a GUT-like symmetry like G(224) or SO(10).

In a theory with RH neutrinos as an essential member of each family,
and with spontaneous breaking of B--L and I$_{3R}$ at a high scale
M$_{\rm B-L}$, both already introduced in Ref.~\cite{ref:1d}, the RH
neutrinos can and generically will acquire a superheavy Majorana
mass ($M(\nu_R) \sim M_{B-L})$ violating B--L by two units. The
$\nu_L$ and $\nu_R$ will of course combine to get a Dirac mass
$(m(\nu)_{\rm Dirac})$ through electroweak symmetry breaking. The
idea of the seesaw~\cite{ref:3a} is simply this. Ignoring family
mixing for a moment, it combines the superheavy Majorana mass of
$\nu_R$ with the Dirac mass (through diagonalization) to yield a
mass for $\nu_L$ given by
\begin{equation}
m(\nu_L) \approx m(\nu)^2_{\rm Dirac}/M(\nu_R) \label{eq:3}
\end{equation}
which is naturally superlight if $M(\nu_R)$ is naturally superheavy.

The seesaw mechanism in turn of course needs the ideas of
SU(4)-color and SUSY unification so as to be quantitatively useful.
This is because these two ideas serve to determine the Dirac and the
Majorana masses (especially for the third family) rather well, which
would otherwise be quite arbitrary.  To be specific, SU(4)-color
treats $\nu^\tau_{R,L}$ as the fourth color partners of $t_{R,L}$.
As a result, it yields (see Eq. 1(b)): $m(\nu^\tau)_{\rm Dirac}
\approx m_{\rm top} (M_{\rm GUT}) \approx 120 \ {\rm GeV} $.  [Here,
we are ignoring family-mixing; which, however, would introduce
insignificant correction to this relation for hierarchical fermion
mass-matrices, see Section 4.]

Furthermore, by having B--L as a local symmetry, SU(4)-color
naturally associates the Majorana mass of $\nu^\tau_R$ with the
(B--L)-breaking scale, and thereby, for the sake of coupling
unification, with the GUT-scale.  In the context of a minimal Higgs
system, which breaks B--L by one unit, one thereby obtains (see next
section): $M(\nu^\tau_R) \sim (M_{B-L})^2/M \approx (M^2_{\rm
GUT}/M) \approx (2 \times 10^{16}\ {\rm GeV})^2/(10^{18}\ {\rm
GeV})(1/2-2)  \approx (4\times 10^{14}\ {\rm GeV})(1/2-2)$. Here
$M_{B-L}$ denotes the VEV of the Higgs field that breaks B--L by one
unit; for reasons mentioned above it is identified with the
GUT-scale. The mass M denotes the scale of an effective
non-renormalizable operator induced by Planck or string-scale
physics (see Section 4); we have therefore set $M \approx (10^{18}\
{\rm GeV})(1/2-2)$. Substituting these values of $m(\nu^\tau)_{\rm
Dirac}$ and $M(\nu^\tau_R)$ into Eq. (\ref{eq:3}) and ignoring 2--3
family mixing for a moment (which turns out to be unimportant for
mass eigenvalues), one thus obtains:
\begin{eqnarray}
m(\nu_{L}^{3}) \approx (120\mathrm{\ GeV})^{2}/(4\times
10^{14}\mathrm{\ GeV}(1/2\mbox{--}2) )\approx (1/28\mathrm{\
eV})(1/2\mbox{--}2))\ .\label{eq:4}
\end{eqnarray}
With hierarchical pattern for fermion mass-matrices (see Sec. 4),
one necessarily obtains $m(\nu_{L}^{2})\ll m(\nu_{L}^{3})$ and thus
$\sqrt{\Delta m_{23}^2}\approx m(\nu_{L}^{3})\sim 1/28 \mathrm{\
eV}(1/2\mbox{--}2)$. This is just the right magnitude matching the
mass scale observed at SuperK~\cite{ref:o}!

\emph{Without an underlying reason as above for at least the
approximate values of these two vastly differing
mass-scales---$m(\nu_{\mathrm{Dirac}}^{\tau})$ and
$M(\nu_{R}^{\tau})$---the seesaw mechanism by itself would have no
clue, quantitatively, to the mass of the LH neutrino.}  In fact it
would yield a rather arbitrary value for $m(\nu_{L}^{\tau}) \approx
m(\nu^3_L)$, which could vary quite easily by more than 10 orders of
magnitude either way around the observed mass scale (that is from
about $10^{-14}$ eV to nearly 10 GeV, say). This would in fact be
true if one introduces the RH neutrinos as a singlet of the SM or of
SU(5).\footnote{To see this, consider for simplicity just the third
family.  Without SU(4)-color, even if a RH two-component fermion $N$
(the analogue of $\nu_R$) is introduced by hand as a \emph{singlet}
of the gauge symmetry of the SM or $SU(5)$, \emph{such an $N$ by no
means should be regarded as a member of the third family, because it
is not linked by a gauge transformation to the other fermions in the
third family}.  Thus its Dirac mass term given by
\(m(\nu_{\mathrm{Dirac}}^{\tau})[ \bar{\nu}_{L}^{\tau} N + h.c. ]\)
can vary from say 100 GeV to even 1~MeV. Likewise, without B--L, the
Majorana mass $M(N)$ can be as high as the Planck or the string
scale (\(10^{18}\mbox{--}10^{17}\mbox{\ GeV}\)), and as low as say
1~TeV.  This would yield $m(\nu^\tau_L)$ varying from about
10$^{-14}$ eV to about 10 GeV. Such arbitrariness both in the Dirac
and in the Majorana masses, is drastically reduced, however, once
$\nu_R$ is related to the other fermions in the family by an
$SU(4)$-color gauge transformation and SUSY unification is assumed
(see discussion preceding Eq. (\ref{eq:4})).} Within symmetries like
G(2213) = SU(2)$_L \times$ SU(2)$_R \times$ U(1)$_{B-L} \times$
SU(3)$^c$~\cite{ref:t} and [SU(3)]$^3$~ \cite{ref:l}, which possess
B--L, RH neutrino is a compelling feature; so the Majorana mass of
$\nu^\tau_R$ can be constrained. But the Dirac mass of $\nu_\tau$ is
not related by any symmetry to the top mass. It can thus vary from
say 1 MeV to 100 GeV.  This would render the prediction of
$\nu^\tau_L$-mass  (within G(2213) or [SU(3)]$^3$) uncertain by
almost ten orders of magnitude for a given $M(\nu^\tau_R)$ (see Eq.
(\ref{eq:3})).  They also do not yield $m_b (M_{\rm GUT}) \approx
m_\tau$ (Eq. 1(a)).  Now, as mentioned before, flip SU(5) $\times$
U(1) possesses $\nu_R$ and B--L, and yields Eq. 1(b), but not Eq.
1(a).\footnote{Although I have argued that the data suggests the
need for SU(4)-color in 4D, I should mention a possible exception.
If either G(2213) or [SU(3)]$^3$ or flip SU(5) $\times$ U(1) emerges
in 4D through a string solution and if the latter turns out to
ensure SU(4)-color relations for the Yukawa couplings (especially
both Eqs. 1(a) and 1(b)) near the string scale, then for all
practical purposes the advantages of SU(4)-color would persist in
4D, since B--L is already contained in the gauge symmetries
mentioned above.}

In short, the seesaw mechanism needs (for a quantitative success)
the ideas of SUSY unification and SU(4)-color, and of course
vice-versa; only \emph{together} they provide an understanding of
neutrino masses as observed, explaining the large hierarchy such as
$m(\nu^\tau_L)/m_{\rm top} \sim 10^{-12}$ (Eq. (\ref{eq:3})).
Schematically, one thus finds:
\begin{equation}
\begin{array}{rcl}
\fbox{\(
\begin{array}{c}
\mbox{SUSY UNIFICATION} \\ \mbox{WITH $SU(4)$-COLOR}
\end{array}\)} & \oplus & \fbox{SEESAW} \\
 & \Downarrow & \\
m(\nu_{L}^{3}) & \sim & 1/10 \mbox{\ eV}.
\end{array}
\label{eq:scheme}
\end{equation}

By the same token, as claimed in the introduction and in Section 2,
the agreement of the theoretically expected $\sqrt{\Delta
m^2(\nu)_{23}}$ with the observed SuperK value, together with the
success of the mass-relations 1(a) and 1(b), clearly seem to favor
the idea of the seesaw and select out the route to higher
unification based on SU(4)-color and supersymmetry, as opposed to
other alternatives.  This is further strengthened by the success of
leptogenesis discussed in Section 6.

\section{\large Fermion Masses and Neutrino Oscillations}

I will now present the results of a specific attempt, by Babu, Pati
and Wilczek~\cite{ref:4a}, to understand some aspects of the masses
and mixings of fermions including the neutrinos in the context of
SO(10) or G(224) symmetry.  For operational purposes, we assume,
together with SO(10), a commuting U(1)-flavor symmetry~\cite{ref:u}
which distinguishes  between different families and leads to the
desired gross hierarchical pattern in the mass-matrices (see below).
Here again one sees how essentially the group theory of SO(10) or
G(224) serves to explain, in the context of a minimal Higgs system
(see below), some intriguing features such as the smallness of
$V_{cb} \approx 0.04$ and the largeness of $\theta^\nu_{23} \approx
\pi/4$ in a compelling manner, together with $m_b({\rm GUT}) \approx
m_\tau$, and $m_s({\rm GUT}) \ne m_\mu$.  Owing to length limitation
I will leave out much of the discussion in this regard, and would
refer the reader to the original paper~\cite{ref:4a}, and also to
the subsequent review talks~ \cite{ref:u}.

The BPW framework assumes that the fermion mass-matrix is generated
through only a {\em minimal Higgs system}, which also serves to
break SO(10) to $SU(3)^c \times U(1)_{em}$.  It consists of the set:
\footnote{Analogous multiplets are used for the case of G(224).}
$H_{\rm minimal} = \left\{ \mathbf{45_H,\ 16_H,\ \overline{16}_H,\
10_H,\ S} \right\} \ , % \label{eq:6}
$ where $S$ is a singlet of SO(10) carrying appropriate U(1)$_F$
flavor-charge~\cite{ref:u}.  This Higgs system is regarded as
minimal in the sense that it consists only of lowest dimensional
multiplets \footnote{Such lowest dimensional Higgs multiplets have
also been considered by other authors~\cite{ref:4b,ref:4c} with
similar degree of success.  For what it is worth, solutions of
weakly interacting heterotic string theories do yield such
low-dimensional multiplets, though not higher dimensional ones like
126 and 120~\cite{ref:abc}.} in contrast to large-dimensional
tensorial multiplets ({\em e.g.} $126_H + \overline{126}_H + 210_H +
{\rm possibly}\ 120_H + 10_H$) which have also been used widely in
the literature~\cite{ref:4d}. The advantages of low-dimensional over
the large-dimensional multiplets and vice verse are discussed
elsewhere~\cite{ref:u} (see, {\em e.g.} the third paper).

Of the Higgs multiplet shown above , the VEV of \(\left\langle
\mathbf{45_{H}} \right\rangle \sim M_{U}\) breaks $SO(10)$ in the
B--L direction to \(G(2213) = SU(2)_{L} \times SU(2)_{R} \times
U(1)_{B-L} \times SU(3)^{c}\), and those of \(\left\langle
\mathbf{16_{H}} \right\rangle = \left\langle
\overline{\mathbf{16}}_{\mathbf{H}} \right\rangle\sim M_U\) along
\(\left\langle \tilde{\bar{\nu}}_{RH} \right\rangle\) and
\(\left\langle \tilde{\nu}_{RH} \right\rangle\)\ break $G(2213)$
into the SM symmetry $G(213)$ at the unification-scale $M_{U}$. Now
$G(213)$ breaks at the electroweak scale by the VEVs of
\(\left\langle \mathbf{10_{H}} \right\rangle\) and of the EW doublet
in \(\left\langle \mathbf{16_{H}} \right\rangle\) to \(SU(3)^{c}
\times U(1)_{em}\).  The singlet $S$ is also assumed to have a VEV
$\sim  M_U$.  It should be noted that although the VEVs of 16$_H$
and $\bar{16}_H$ break B--L by one unit and thereby break the
familiar R-parity, the system still preserves a discrete matter
parity ($16_i \rightarrow -16_i$, $16_H \rightarrow 16_H$,
$\bar{16}_H \rightarrow \bar{16}_H$, $45_H \rightarrow 45_H$, $10_H
\rightarrow 10_H$, $S\rightarrow S$, gauge multiplet $\rightarrow$
itself).  This suffices to ensure a stable LSP and absence of d = 4
proton decay operators.

With the minimal Higgs system, fermions receive Dirac masses through
three types of allowed SO(10)-invariant effective couplings: (i)
$h_{ij}\, 16_i\, 16_j\, 10_H$; \break (ii) $a_{ij}\, 16_i\, 16_j\,
10_H\, 45_H/M$; and (iii) $g_{ij}\, 16_i\, 16_j\, 16_H\, 16^d_H/M$,
where $(i,j) = 1,2,3$ represent family indices.\footnote{Here
16$^d_H$ denotes that the lepton-like doublet in 16$_H$, having the
quantum numbers of H$_d$.  It acquires a VEV of the electroweak
scale due to its mixing with the down-type doublet in
10$_H$.~\cite{ref:4a}} Each of these coupling parameters like
$h_{ij},\, a_{ij}$ and $g_{ij}$ carry suitable powers of $(\langle
S\rangle/M)$, depending upon flavor symmetry~\cite{ref:u}.  Such
powers and/or higher dimensional operators like (ii) and (iii)
induce the desired hierarchical couplings with ``33" $\gg$ ``32"
$\sim$ ``23" $\gg$ ``22" $\gg$ ``12", etc.\footnote{As an example,
the desired hierarchical pattern mentioned above can be achieved
simply by assigning the U(1)$_F$-charges as follows: (a, a+1, a+2,
-2a, -a-1/2, -a,0,-1) to (16$_3$, 16$_2$, 16$_1$, 10$_H$, 16$_H$,
$\overline{16}_H$, 45$_H$, S), with a = 1/2.  Origin of such a
flavor symmetry must be sought in an underlying theory like string
theory.} The coupling $h_{33}\, 16_3\, 16_3\, 10_H$, allowed by
SO(10) and U(1)$_F$ flavor symmetry, is the leading term with
$h_{33} \approx h_{\rm top} \sim 1$, that gives masses to the third
family. Higher dimensional operators like (ii) and (iii) are
expected to arise through quantum gravity or string-scale physics so
that $M\sim M_{\rm st}$ or $M_{\rm pl} \sim 10^{18}$
GeV(1/2-2).\footnote{I should add that some authors have exhibited a
strong preference for using only renormalizable Yukawa interactions
(effective or not). Such a preference does not seem to be warranted,
however, because effective (apparently) non-renormalizable
interactions do routinely arise at low energies by utilizing purely
renormalizable interactions at high energies. That is the case for
example for the four fermion weak interaction, and also for the
effective interactions leading to seesaw neutrino masses (Type I or
Type II). In the present case, one knows that there exists physics
above the GUT scale, characterized by the string or the Planck
scale.  Thus, if one ignores such effective interactions, which are
allowed by all symmetries, and which would be relevant especially
for the masses of the first two families, the question might arise:
Why?} The coupling involving $\VEV{10_H}\cdot \VEV{45_H}/M$ provides
(B--L)-dependent family-antisymmetric coupling that contributes only
to off-diagonal mixings, while $\VEV{16_H}\cdot \VEV{16^d_H}/M$
contributes only to the down flavor sector and thereby to CKM
mixings~\cite{ref:4a}.

In addition to these couplings, which provide Dirac masses, the RH
neutrinos receive hierarchical Majorana masses through effective
couplings of the form $f_{ij}\, 16_i\, 16_j$\, $\overline{16}_H$\,
$\overline{16}_H/M$; where $f_{ij}$'s contain appropriate powers of
$\langle S\rangle/M$, dictated by flavor symmetry~\cite{ref:u}. The
leading contribution arises again for the third family with $f_{33}
\sim 1$, so that $M(\nu^3_R) \approx f_{33}
\langle\overline{16}_H\rangle^2/M \approx (2\times 10^{16}\ {\rm
GeV})^2/(10^{18}\ {\rm GeV})(1/2-2) \approx 4 \times 10^{14} \ {\rm
GeV}(1/2-2)$ (as noted in Section 3).

Assuming for simplicity CP conservation for a moment~\cite{ref:4a},
so that all entries in the fermion mass-matrices are real, it turns
out that the number of effective parameters (7 for the Dirac sector)
are significantly less than the observables (12 for the quark and
charged lepton sectors).  We determine these parameters by using
$m^{\rm phys}_t = $ 174 GeV, $m_c(m_c) =$ 1.37 GeV, $m_s$ (1 GeV) =
110--116 MeV, $m_u$ (1 GeV) = 6 MeV and the masses of $e$, $\mu$ and
$\tau$ as inputs.  For the Majorana mass matrix of the RH neutrinos,
U(1)-flavor symmetry~\cite{ref:u} fixes the ``23" element $y$ to be
of order 1/10 relative to the ``33" element, which is normalized to
1 in units of $M(\nu^3_R)$.  It turns out~\cite{ref:4a} that with
such a hierarchical $|y|\sim 1/10$, an input value of
$m(\nu_2)/m(\nu_3)\approx 1/5-1/7$ (as suggested by the data) can be
satisfied provided $y$ is {\em negative} \footnote{In turn, this
negative sign for $y$ implies~\cite{ref:4a} that the contributions
to $\theta^\nu_{23}$ from the charged lepton and neutrino sectors
{\em add to each other} rather than subtract (see Eq. (6)). Equally
important is the fact that the charged lepton contribution to
$\theta^\nu_{23}$ gets enhanced relative to the familiar mixing
angle of $\sqrt{m_\mu/m_\tau}$ by a (B--L)-dependent
family-antisymmetric factor
$(\eta-3\epsilon)^{1/2}/(\eta+3\epsilon)^{1/2}$, while the analog of
this last factor provides a suppression for $V_{cb}$, just as
needed! (See the references~\cite{ref:4a,ref:u} for details of this
discussion.)  The combination of these two ingredients ({\em i.e.}
addition of the two contributions and (B--L)-dependent enhancement)
ends up in yielding a nearly maximal $\theta^\nu_{23}$, while
$V_{cb}$ is small.} with a value $y \approx -1/17$. One is then led
to the following seven predictions:

\bigskip
\eject

 \hbox{{\bf \hspace{1in}Predictions\hspace{1.75in}
Observations}}
\medskip
 \begin{tabular}[t]{rll}
$m_b(m_b)$ &$\ \approx\quad $(4.7--4.9) GeV
&$\ \approx\quad $ 4.2 GeV\\[1ex]
$\sqrt{\Delta m^2(\nu)_{23}}$ &$\ \approx\quad $ (1/24 eV)(1/2--2)
& $\ \approx\quad $ 1/20 eV\\[1ex]
$V_{cb}$ &$\ \approx\quad $  0.044
&$\ \approx\quad $ 0.042 \\[1ex]
$\theta^\nu_{23}$
 & $\ \approx\quad $$ \left|(0.437)_{ch.lep.} + \sqrt{m(\nu_2)/m(\nu_3)} \
\right|$
&  \\[-3ex]
\raisebox{2.5ex}{$\Bigg\{$} $\sin^22\theta^\nu_{23}$ &$\
\approx\quad $$ 0.97-0.995 \left({\rm for}\
\frac{m(\nu_2)}{m(\nu_3)}\approx \frac{1}{7}-\frac{1}{5}\right)$
&$\ \approx\quad $ 0.90--1 \\
$V_{us}$ &$\ \approx\quad $ 0.20
&$\ \approx\quad $ 0.225\\[1ex]
$V_{ub}$ &$\ \approx\quad $ 0.0031
&$\ \approx\quad $ 0.0039    \\[1ex]
$m_d$ (1 GeV) &$\ \approx\quad $ 8 MeV & $\ \approx\quad $5--8 MeV\
. ~~~~ (7)
\end{tabular}
\setcounter{equation}{7}
\bigskip

Leaving aside $|V_{ub}|$, which is lower than the observed value by
about 20\%,\footnote{Such 20\%\ discrepancy in $|V_{ub}|$ gets
corrected without affecting other entries  when one includes CP
violation (see next section).} it is indeed remarkable, considering
the bizarre pattern of fermion masses and mixings, that all the
remaining six predictions agree with the data to within about 10\%!
Particularly intriguing are the (B--L)-dependent enhancement for
$\theta^\nu_{23}$ versus suppression for $V_{cb}$, as well as the
fact that the contributions from the charged lepton and neutrino
sectors necessarily add~\cite{ref:4a} in the case of
$\theta^\nu_{23}$ (for a hierarchical $y$) (see Eq. (7)) but the
corresponding terms for $V_{cb}$ subtract.   {\em It is the
combination of these two factors which end up in yielding a nearly
maximal $\theta^\nu_{23}$, and simultaneously a small $V_{cb}$, as
needed (see footnote m  and references~\cite{ref:4a,ref:u})}. It
should be stressed that {\em this explanation of the largeness of
$\theta^\nu_{23}$ is characteristically different from all the other
explanations in the literature}.  (Contrast for example from the
lop-sided SO(10)-models~\cite{ref:4b}, where the largeness of
$\theta^\nu_{23}$ arises entirely from the charged lepton sector.)
It should also be noted that although $m(\nu_2)/m(\nu_3)$ varies
with $y$, the prediction of $\theta^\nu_{23}$ is fairly stable as
long as $|y|$ has a hierarchical value ($\sim 1/10$, within a factor
of 3, say), as suggested by flavor symmetry~ \cite{ref:u}. In this
sense, the near maximality of $\theta^\nu_{23}$ is a {\em compelling
prediction} of the model.

It has been noted~\cite{ref:u} that small non-seesaw contribution
arising from a higher dimensional operator, consistent with flavor
symmetry and SO(10), can contribute to the $\nu^e_L\nu^\mu_L$ mass
term $\sim$ few $\times 10^{-3}$ eV;  this can quite plausibly lead
to large $\nu_e-\nu_\mu$ oscillation angle in accord with the LMA
MSW solution to the solar neutrino puzzle:  Including this we get:
\begin{eqnarray}
\label{eq:nu12}
\begin{array}{l}
\ \ \ m(\nu_2)\approx (9-6)\times 10^{-3} \ {\rm eV\ (from\
seesaw)}~~~~~~~~~~~~~~~~~~~~~~~~~~~~~~~~~~~~~~~~(a)\\[1ex]
\begin{array}{l}
\ \ m(\nu_1)\approx (1-{\rm few})\times 10^{-3}\ {\rm eV};  {\rm
thus\ \Delta
m^2_{12}\approx (4-8)\times 10^{-5} eV^2} ~~~~~~~~~~~~~ (b)\\[1ex]
\begin{array}{l}
\ \sin^2 2\theta^\nu_{12}\approx (0.5-0.7)\quad {\rm
(from\ non\ seesaw)}~~~~~~~~~~~~~~~~~~~~~~~~~~~~~~~~~~~~~~ (c)\\[1ex]
\begin{array}{l}
\theta_{13}\lsim (2-5)\times 10^{-2}\ .\quad ~~~~~~~~~~~~~~~~~~~~~~~~~~~~~~~~~~~~~~~~~~~~~~~~~~~~~~~~~~~~~~(d)\\
\end{array}
\end{array}
\end{array}
\end{array}
\end{eqnarray}
Here, $m(\nu_2)/m(\nu_3)$ and thus $m(\nu_2)$ is an input that fixes
the choice of $y$ (see above).   It should be mentioned that in
contrast to the near maximality of $\theta^\nu_{23}$, which emerges
as a compelling prediction, largeness of $\theta^\nu_{12}$ is a
plausible possibility, but not a prediction of the framework. The
superheavy Majorana masses of the RH neutrinos $\nu^i_R$ also get
fairly fixed within the model by the flavor-hierarchy, one
gets~\cite{ref:4a,ref:u}:
\begin{equation}
(M_3,\,M_2\,,M_1) \approx\left(4 \times 10^{14} (1/2-2);\ 10^{12}
(1/2-2);\  4 \times 10^{10} (1/8-4)\right)\, {\rm GeV} \ .
\end{equation}
Note that both the light and the superheavy neutrino masses have a
normal hierarchy: $m_1 \lsim m_2 \ll m_3$ and $M_1 \ll M_2 \ll M_3$.
The latter hierarchy would be important for leptogenesis.

\section{\large CP and Flavor Violations}

To incorporate CP violations into the G(224)/SO(10)-framework
outlined in the previous section~\cite{ref:4a}, the simplest and
most attractive possibility appears to be to assume that CP
violation arises spontaneously, either entirely or dominantly,
through the VEVs of some or all of the Higgs fields (in our case
45$_H$, 16$_H$, $\overline{16}_H$, 10$_H$ and/or $S$), and thus
through phases in the fermion mass matrices~\cite{ref:5a}.  These
latter would induce CP violation both through SM CKM-interactions as
well as through supersymmetric interactions involving
sfermion-gaugino loops.

Motivated by the feebleness of flavor changing neutral current
interactions, we assume that SUSY-breaking is flavor-universal and
CP-conserving at a high scale ($M^* \gsim M_{\rm GUT}$).  SUSY
contributions still violate CP and flavor owing to renormalization
group-evolution of the scalar masses and the A-parameters from the
high messenger scale $M^*$ to low scales, involving Yukawa couplings
which are flavor-dependent~\cite{ref:5aa}, and because of phases in
the fermion mass matrices.

The challenge then is this:  Can the idea outlined above be
implemented by including both the SO(10)-based SM and the SUSY
contributions so that the framework would be consistent with the
{\em observed} CP and/or flavor violations---as in $\Delta m_K$,
$\epsilon_K$, $\Delta m_{B_d}$ and $S(B_d \rightarrow J/\psi\,
K_S$)---while still preserving its successes~\cite{ref:4a} listed in
the previous section as regards the predictions of the fermion
masses and neutrino oscillations (see Eq. (6))?  Furthermore, does
the SUSY G(224)/SO(10)-framework with CP violation introduced as
above~\cite{ref:5a} lead to some characteristic predictions that can
help distinguish it from the SM and other alternative frameworks?

The first question raised above poses a non-trivial challenge
especially because {\em all four CP and/or flavor violating entities
listed above agree rather well with the predictions of the standard
CKM model for a single choice of the Wolfenstein parameters (see
{\em e.g.}  Ref.~\cite{ref:5b}):}
\begin{equation}
\bar\rho_W = 0.204 \pm 0.035\, : \qquad \bar\eta_W = 0.336 \pm
0.021\ (1 \sigma)
 \ . \label{eq:5a}
\end{equation}

The questions raised above are addressed in three recent papers by
Babu, Parul Rastogi and myself~\cite{ref:5a,ref:5c,ref:5d}.  It
turns out that the answers to both questions are in the affirmative.
True distinctions of the SUSY G(224)/SO(10) framework~\cite{ref:4a}
outlined above from the SM as well as alternative SO(10)-frameworks
({\em e.g.} of Albright-Barr~\cite{ref:4b}) arise though its
predictions, for example, for the edm's of the neutron and the
electron and the rates for $\mu \rightarrow e \gamma$ and $\tau
\rightarrow \mu\gamma$.  I will leave out the discussions in this
regard and mention only the essence of the results.  I would refer
the reader to the original papers~\cite{ref:5a,ref:5c,ref:5d}  for
details of clarification.  The results are as follows:

\begin{enumerate}
\item
Allowing for phases ($\sim$ 1/10 to 1/2) in the parameters of the
G(224)/SO(10)-based fermion mass matrices~\cite{ref:4a}, it is found
that there do exist solutions which yield masses and magnitudes of
mixings of quarks and leptons including neutrinos, all in good
accord with observations (to within 10\%), and at the same time
yield the following values for the Wolfenstein parameters:
\begin{equation}
(\bar\rho_W)_{SO(10)} \approx 0.14 - 0.17 \quad {\rm and}\quad
(\bar\eta_W)_{SO(10)} \approx 0.35 - 0.38 \ . \label{eq:5b}
\end{equation}
Note that these are fairly close to the SM-based phenomenological
values (Eq.~(\ref{eq:5a})).  This, together with having the right
magnitudes of the CKM mixings, is the reason why the SUSY
G(224)/SO(10) model as developed in Refs.~\cite{ref:4a,ref:5a}
succeeds in accounting for the observed CP and flavor violation in
the quark sector (see below).\footnote{Note that a priori a given
SO(10) model with a certain pattern of fermion mass matrices need
not yield ($\bar \rho_W, \bar\eta_W$) lying even in the first
quadrant (not to mention having the right magnitudes), for {\em any
choice} of phases and magnitudes of the parameters of the
mass-matrices, without conflicting with the observed fermion masses
and mixings (see {\em e.g.} Ref.~\cite{ref:bbb}).}

\item
{\bf CP and Flavor Violations in Quark Sector}.  Including both the
SO(10)-based SM-contribution (and thus using Eq. (\ref{eq:5b})) and
the SUSY-contribution\footnote{I should stress that for a given
choice of the few SUSY parameters ({\em i.e.} $m_0$, $M_{1/2}$ and
$M^*$), SUSY CP and flavor violations (both the magnitudes and the
phases) are completely determined in the model owing to prior
works~\cite{ref:4a,ref:5a} that successfully describes fermion
masses and neutrino oscillations and yields $(\bar\rho_W, \eta_{\bar
W})$ as in Eq. (\ref{eq:5b}).} (with a plausible choice of the
SUSY-spectrum---{\em e.g.} corresponding to $m_o \sim$ (550--650)
GeV and $m_{1/2} \sim$ (250--300) GeV), $m_{sq} \approx (0.8-1)$
TeV, $m_{\tilde \ell} \sim$ 600 GeV, $m_{\tilde g} \sim$ 120 GeV,
and $x = m_{\tilde g}/m^2_{sq} \approx$ 0.6--0.8, $M^*/M_{\rm GUT}
\approx 3$ we obtain~\cite{ref:5a}:
\begin{eqnarray}
&&(\Delta m_K)_{\rm short\ dist} \approx 3\times 10^{-15}\ {\rm
GeV}\, ;\quad \epsilon_K \approx (2-2.5)\times 10^{-3}\nonumber\\
&&(\Delta m_{B_d}) \approx (3.5-3.6) \times 10^{-13}\, {\rm GeV}\, ;
\quad S(B_d\rightarrow J/\psi\, K_s) \approx 0.68-0.74\nonumber \\
&&S(B_d\rightarrow \phi K_s) \approx 0.65 - 0.73 \nonumber \\
&&\Delta m_{B_s} \approx 17.3\ ps^{-1}\ ;\quad S(B_s\rightarrow
J/\psi\, \phi) \approx {\rm few} \ \% \nonumber \\
&&A(b\rightarrow s\gamma)^{SO(10)}_{SUSY} \approx (1-5)\%\ {\rm of}
\ A(b\rightarrow s\gamma)_{SM} \nonumber \\
&&(edm)^{\rm neutron}_{Aind} \approx (1.6,\, 1.08) \times 10^{-26}\,
ecm \ (\tan\beta=5,10) \nonumber \\
&&(edm)^{\rm electron}_{Aind} \approx (1.1 \times
10^{-28}/\tan\beta)\, ecm \ .
\end{eqnarray}
Here, we have used central values of the lattice results on the
matrix element $\widehat B_k = 0.86$, $f_{B_d} \sqrt{\widehat
B_{B_d}}$  = 215 MeV, and $f_{B_s} \sqrt{\widehat B_{B_s}}$ = 245
MeV.  Note that the first four entities ($\Delta m_K,\ \epsilon_K,\
\Delta m_{B_d}$ and $S(B_d \rightarrow J/\psi\, K_s)$), which are
well measured, are all in good agreement with observations (within
10\%).  In all these cases, the SUSY-contribution turns out to be
rather small ($\leq$ 5\%\ in amplitude) compared to the
SM-contribution, except however for $\epsilon_K$, for which it is
sizeable ($\approx$ 20--30\%) and relatively {\em negative}.  This
negative contribution to $\epsilon_K$ can help distinguish the SUSY
G(224)/SO(10) model from the SM, once the matrix element $\widehat
B_K$ is determined to 5\%\ accuracy.  Given the present experimental
limit on $(edm)_n < 6.3 \times 10^{-26}\ ecm$, the predicted value
of $(1.6-1)\times 10^{-26}\, ecm$ is in an extremely interesting
range and can be probed with an improvement in the current limit by
a factor of 10.  The predicted value of $S(B_d\rightarrow \phi
K_s)$, which is close to the SM-value but is at present about
2$\sigma$ away from the BaBar-BELLE average value of (0.47 $\pm$
0.19), would also provide a very interesting test once it is better
measured, and so also would the parameters of the $B_s$ system,
which are predicted to nearly coincide with the predictions of the
SM.
\end{enumerate}

As a summary of this section, {\em we see that the SUSY G(224) or
SO(10)-framework (remarkably enough) has met all the challenges so
far in being able to reproduce the observed features of  CP and
quark-flavor violations, as well as of fermion masses and neutrino
oscillations; and it has predictions than can probe the framework
further}!

\section{\large Lepton Flavor Violation}

Lepton flavor violating processes (such as $\mu \rightarrow
e\gamma$, $\tau \rightarrow \mu\gamma$, $\mu N\rightarrow eN$, etc.)
can provide sensitive probes into SUSY grand unification.  In our
case, these get contributions from three sources~\cite{ref:5c}: (i)
Slepton (mass)$^2$-elements $(\delta m^2)^{ij}_{LL}$ arising from
RG-running of scalar masses from $M^*$ to $M_{\rm GUT}$ ($M^*$ being
the messenger scale where SUSY-breaking parameters are
flavor-universal); (ii) $(\delta m^2)^{ij}_{LR}$ which arise from
A-terms induced through RG-running from $M^*$ to $M_{\rm GUT}$; and
(iii) $(\delta m^2)^{ij}_{LL}$ arising from RG-running of scalar
masses from $M_{\rm GUT}$ to the RH neutrino masses $M_{Ri}$.  The
first two effects, though important as long as $\ell n(M^*/M_{\rm
GUT}) \gsim 1$, are invariably dropped in the literature.  For a
given choice of the SUSY spectrum ({\em i.e.} $m_0$ and $m_{1/2}$)
and $M^*$, {\em the predictions of the model are completely fixed in
terms of the mass-matrices of the quarks, leptons and neutrinos,
which as we saw are fixed following discussions in Sections 4 and
5~\cite{ref:4a,ref:5a}.}

The predictions for the branching ratios for the SO(10)-model (for a
sample choice of $(m_0$, $m_{1/2}$), with $\ell n\, M^*/M_{\rm GUT}
=1$; {\em i.e.} $M^*\approx 2.7\ M_{\rm GUT}$) are as
follows~\cite{ref:5c}:

\bigskip
{\footnotesize  %\hoffset-.3in
 \noindent
\begin{tabular}{c|c|c|c|c}
($m_0,m_{1/2}$) & (600, 300) & (450, 300) & (100, 440) & (400,
300)\\[1ex]
$B(\mu \rightarrow e\gamma)$  & $(3.3,\, 9.8) \times 10^{-12}$ &
$(2.7,4.6)\times 10^{-11}$ & $(1,1) \times 10^{-8}$ &
$(0.95,3.8)\times 10^{-11}$ \\[1ex]
$B(\tau\rightarrow\mu\gamma)$& $(2.4,3.1)\times 10^{-9}$ &
$(2.7,5.6)\times 10^{-9}$ & $(8.3,8.4)\times 10^{-8}$&
$(1.4,1.8)\times 10^{-8}$\\
\end{tabular}}
\bigskip

\noindent
 Here, $(m_0,m_{1/2})$ are given in GeV.  The two entries
appearing under each column correspond to $\mu > 0$ and $\mu < 0$
respectively. The first three columns are for $\tan\beta$ = 10 and
the last one for $\tan\beta$ = 20.  {\em The predictions for
$B(\mu\rightarrow e\gamma)$ for the G(224)-model are lower by a
factor $\approx$ 4--6 than those for the SO(10)-model~\footnote{This
is because the RG running of the scalar masses and the A-parameters
from $M^*$ to $M_{\rm GUT}$ leads to larger effects for the case of
SO(10) than for G(224).  For distinguishing between these models,
one would need to determine $\tan\beta$.} (see
Ref.~\cite{ref:5a,ref:5c}).} Given the present empirical limit of
$B(\mu \rightarrow e\gamma) \leq 1.2 \times 10^{-11}$~\cite{ref:6a},
we see that the case of $(m_0,\, m_{1/2})$ = (100, 440) GeV, is
excluded for the SO(10) as well as for the G(224) model, while
$(m_0,\, m_{1/2})$ = (450, 300) GeV is excluded for the former
though not for the latter, with $\tan\beta$ = 10. The interesting
point is that, even if sleptons are rather heavy (($m_0,\, m_{1/2})
\approx$ (800, 250) GeV, say), for which one finds~\cite{ref:5c}
($B(\mu \rightarrow e\gamma))_{SO(10)} \approx$ (2.9, 17) $\times
10^{-13}$ for $(\mu > 0,\ \mu < 0)$, $\mu \rightarrow e \gamma$ {\em
should be discovered with an improvement in the current limit by a
factor of 10--100}. This is being planned at the MEG experiment at
PSI. The present empirical limit of $B(\tau \rightarrow \mu\gamma) <
7 \times 10^{-8}$ obtained at BaBar~\cite{ref:6b} also excludes the
choice $(m_0,\, m_{1/2})$ = (100, 440) Gev for the case of SO(10)
with $\tan\beta$ = 10.  One finds~\cite{ref:5c}, however, that even
if sleptons are rather heavy $(\lsim$ 800 GeV, say), $\tau
\rightarrow \mu\gamma$ should be discovered with improvement in the
current limit by a factor of 10--50.

Thus, we see that studies of lepton flavor violation can provide
stringent tests of the SUSY SO(10)/G(224)-framework.  They can even
distinguish between SO(10) and G(224) symmetries.  Although, I have
not explicitly discussed, it turns out that they can also clearly
distinguish between (for example) the hierarchical~\cite{ref:4a}
versus lop-sided~\cite{ref:4b} SO(10)-models (see
Ref.~\cite{ref:5d}).

\section{\large Baryogenesis Via Leptogenesis}

The observed matter-antimatter asymmetry is an important clue to
physics at truly short distances.  Given the existence of the RH
neutrinos~\cite{ref:1d}, which naturally can possess superheavy
Majorana masses, violating lepton number by two units, baryogenesis
via leptogenesis~\cite{ref:r} has emerged as perhaps the most viable
mechanism for generating the observed baryon asymmetry of the
universe.  The intriguing feature is that it relates our
understanding of the neutrino masses to our own origin.

The question of whether this mechanism can quantitatively explain
the observed baryon asymmetry depends however crucially on the Dirac
as well as the Majorana mass-matrices of the neutrinos, including
the phases therein and the eigenvalues of the Majorana matrix ($M_1,
M_2$ and $M_3$).  This question has been considered in a recent
work~\cite{ref:s} based on a realistic
G(224)/SO(10)-framework~\cite{ref:4a,ref:5a} for fermion masses,
neutrino oscillations and CP violation, as described in Sections 4
and 5.  The advantage in this case is that the Dirac and the
Majorana mass matrices including the phases in the Dirac sector are
already determined by prior considerations~\cite{ref:4a,ref:5a}.
This work has also been reviewed in recent talks~\cite{ref:u}. Here,
I will present only the results and refer the reader to the
references~\cite{ref:s,ref:u} for details.

The basic picture is this.  Within an inflationary scenario, the
lightest RH neutrinos $(N_1$'s) with a mass $\approx 10^{10}$ GeV
($\frac{1}{3}$--3) are produced either from the thermal bath
following reheating ($T_{RH}\approx$ few $\times 10^9$ GeV), or
non-thermally from the decay of the inflaton (with $T_{RH}$ in this
case being about $10^7$ GeV).  In either case, the RH neutrinos
$(N_1$'s) having Majorana masses decay by utilizing their Dirac
couplings into both $\ell + H$ and $\bar\ell + \bar H$ (and
corresponding SUSY modes), thus violating B--L.  In the presence of
CP violating phases, these decays produce a net lepton-asymmetry
$Y_L = (n_L-n_{\bar L})/s$ which is converted to a baryon asymmetry
$Y_B = (n_B - n_{\bar B})/s = CY_L\ (C \approx -\frac{1}{3}$ for
MSSM) by the electroweak sphaleron effects.  For the thermal case,
using $(M_1/M_2) \approx 4 \times 10^{-3}$ (see Eq. (9)), one
obtains~\cite{ref:s}:
\begin{equation}
(Y_B)_{\rm Thermal}/\sin 2\phi_{\rm eff} \approx (10-30) \times
10^{-11} \ . \label{eq:7a}
\end{equation}
Here $\phi_{\rm eff}$ denotes an effective phase depending upon
phases in the Dirac as well as Majorana mass-matrices of the
neutrinos. For the non-thermal case, using an effective
superpotential so as to implement hybrid inflation~\cite{ref:7a}
involving the GUT-scale VEVs of $(1, 2, 4)_H$ and $(1,2,\bar
4)_H$,and $M_1\approx (2\times 10^{10}$ GeV) (1-1/3) in accord with
Eq. (9), one obtains~\cite{ref:s}
\begin{equation}
\label{eq:7b} (Y_B)_{\rm Non-thermal}/\sin 2\phi_{\rm eff} \approx
(20-200)\times 10^{-11} \ ,
\end{equation}
with a reheat temperature $\approx (3-1)\times 10^7$ GeV, in accord
with the gravitino-constraint.  We see that the derived values of
$Y_B$ can in fact account for the recently observed value
$(Y_B)_{\rm WMAP} \approx (8.7\pm 0.4) \times 10^{-11}$, for very
natural values of the phase angle---that is $\sin 2\phi_{\rm eff}
\approx$ (1/3-1) for the thermal case and $\sin2\phi_{\rm eff}
\approx$ (1/2 -- 1/20), for the non-thermal case.

In summary, we see that the SUSY
G(224)/SO(10)-framework~\cite{ref:4a,ref:5a} described in Sections 4
and 5 provides {\em a simple and unified description} of not only
fermion masses, neutrino oscillations, {\em and} CP violation, but
also of baryogenesis via leptogenesis.  In here, we see a beautiful
link between our understanding of the light neutrino masses (using
seesaw, SU(4)-color and SUSY unification) and our own origin!

\section{\large Proton Decay}

Proton Decay is perhaps the most dramatic prediction of grand
unification possessing quark-lepton
unification~\cite{ref:1c}-\cite{ref:1e}.  I have discussed proton
decay in the context of the SUSY SO(10)/G(224)-framework, as
presented here, in some detail in recent reviews~\cite{ref:u}, which
are updates of the results of Ref.~\cite{ref:4a} and other works.
Here I will present only a summary of the results.

In SUSY unification there are in general {\em three distinct
mechanisms} for proton decay: (i) {\bf The familiar d=6 operators}
mediated by $X$ and $Y$-type gauge bosons of SU(5) and SO(10), which
lead to $e^+\pi^0$ as the dominant mode with a lifetime $\sim
10^{35\pm1}$ years; (ii) {\bf the ``standard" d=5
operators}~\cite{ref:8a}, which arise through the exchange of
color-triplet Higgsinos which are in $5_H+\bar 5_H$ of SU(5) or
$10_H$ of SO(10); these lead to $\bar\nu K^+$ and $\bar\nu\pi^+$ as
the dominant modes with lifetimes varying from about $10^{29}$ to
$10^{34}$ years; and (iii) {\bf the ``new" neutrino mass-related d=5
operators} which can generically arise through the exchange of
color-triplet Higgsinos in the $(16_H + \bar{16}_H)$ of
SO(10)~\cite{ref:4a,ref:8b}; these also lead to $\bar\nu K^+$ and
$\bar\nu\pi^+$ as the dominant modes with lifetimes quite plausibly
in the range of $10^{32}-10^{34}$ years.  {\em One important feature
of these ``new" d=5 operators is that they can contribute to proton
decay for either SO(10) or an effective G(224)-symmetry, and for the
latter, they are the only source of proton decay.}

Guided by recent calculation based on quenched lattice QCD in the
continuum limit~\cite{ref:8c} and that of renormalization factors
$A_L$ and $A_S$ for d=5~\cite{ref:8d}, we take (see
Ref.~\cite{ref:u} for details): $|\beta_H|\approx |\alpha_H| \approx
(0.009\ {\rm GeV}^3)(1/\sqrt 2 - \sqrt 2)$; $m_{\tilde q} \sim
m_{\tilde \ell} \sim$ 1.2 TeV (1/2--2); $m_{\widetilde W}/m_{\tilde
q} \simeq$ 1/6\, (1/2--2); $A_L \approx 0.32, \ A_S \approx 0.93;\
\tan\beta \approx 3; \ M_X \approx M_Y \approx 10^{16}$ GeV (1 $\pm$
25\%); and $A_R(d=6,\ e^+\pi^0) \approx 3.4$. Updating the detailed
analysis of Ref. \cite{ref:4a}, the  predicted rates of proton
decay, for the SUSY SO(10)/G(224)-framework~\cite{ref:4a}, with the
parameters as given above, have been presented in some detail in my
review talks~\cite{ref:u}.  (Analogous studies have also been
carried out by the authors of Ref.~\cite{ref:8e} and also by other
authors; see Ref.~\cite{ref:u}.)

As a summary, with the inclusion of the standard as well as the
``new" neutrino-mass related d = 5 operators, one obtains as a
conservative upper limit:~\cite{ref:u,ref:4a}
\begin{equation}
\Gamma^{-1}_{\rm proton} (d=5) \leq \left(\frac{1}{3}-2\right)
\times 10^{34}\ {\rm years} \ \left({\rm SUSY\atop
SO(10)/G(224)}\right) \label{eq:8a}
\end{equation}
with $\bar \nu K^+$ and $\bar \nu \pi^+$ as the dominant modes, and
quite possibly $\mu^+K^0$ being prominent.  For the d = 6,
$e^+\pi^0$-mode, with parameters as mentioned above, one obtains:
\begin{equation}
\Gamma^{-1}_{d=6} (p\ \rightarrow e^+\pi^0)_{\rm Theory} \approx
10^{35\pm 1}\ {\rm years} \ . \label{eq:8b}
\end{equation}
These should be compared with the experimental limits set by superK
studies~\cite{ref:8f}:
\begin{eqnarray}
\Gamma^{-1} (p\ \rightarrow e^+\pi^0)_{\rm expt} &\geq& 6\times
10^{33}\ {\rm years} \nonumber \\
\Gamma^{-1}(p \rightarrow \bar\nu K^+)_{\rm expt} &\geq& 2\times
10^{33}\ {\rm years} \ . \label{eq:8c}
\end{eqnarray}
The implications of the theoretical predictions vis a vis the
present experimental limits on a next-generation detector are noted
in the next section.

\section{\large A Summary}

In this talk, I have argued on empirical grounds that the evidence
in favor of SUSY grand unification possessing the symmetry
SU(4)-color in 4D is rather strong.  It includes:

$\bullet$ Quantum numbers of the members of a family,

$\bullet$ Quantization of electric charge,

$\bullet$ $Q_{e^-}/Q_p = -1$,

$\bullet$ Gauge coupling unification,

$\bullet$ $m^0_b \approx m^0_\tau$,

$\bullet$ $\sqrt{\Delta m^2(\nu)_{23}} \approx 1/20\ eV$,

$\bullet$ A maximal $\theta^\nu_{23} \approx \pi/4$ with a minimal
$V_{cb}\approx$ 0.04, and

$\bullet$ Baryon Excess $Y_B\sim 10^{-10}$.

\noindent All of these features and more including (even) CP and
flavor violations hang together neatly {\em within a single unified
framework} based on a presumed string-derived G(224) or
SO(10)-symmetry in 4D, with low-energy supersymmetry.  It is hard to
imagine that the neat fitting of {\em all these pieces} can be a
mere coincidence.  It thus seems pressing that dedicated searches be
made for the two main missing pieces of this picture---that is
supersymmetry and proton decay.  The search for supersymmetry at the
LHC and a possible future ILC is eagerly awaited.  That for proton
decay would need an improvement in the current sensitivity by about
a factor ten (see previous section). That in turn will need a
megaton-size underground detector which could also probe deeper into
neutrino-physics as well as supernova physics.  The need for such a
detector can thus hardly be overstated.

\section{\large Concluding Remarks:  A Wish and a Goal}

Before ending this talk I would like to say a few words on a point
of view that I have alluded to in the introduction and have
implicitly adopted throughout in the rest of the talk.  This has to
do with the presumed origin of a desired effective theory in 4D near
the GUT/Planck scale from an underlying theory.

I have argued {\em on empirical grounds} that there is a  need for
having an effective grand unification-like symmetry possessing
G(224) or SO(10)-symmetry in 4D.  Clearly such a symmetry aids much
to our understanding as evidenced by the list in the preceding
section. It, however, leaves much to be explained by a deeper
theory. Needless to say, the origin of the three families, their
hierarchical masses and mixings,  the values of certain parameters
(such as $\alpha_{em}$, $\alpha_{st}$, $G_F$, $m_e$, $m_e/m_p$,
$m_n-m_p$ and $G_N$),  the fact that we live in four dimensions, and
most of all the utterly minuscule magnitude of the cosmological
constant (dark energy), many of which appear to be ``chosen''
somehow so as to satisfy anthropic constraints, are among the
deepest puzzles confronting us today.

I have also implicitly assumed the existence of an underlying
unified theory including gravity---be it string/M theory or
something yet unknown---that would describe nature in a predictive
manner and explain some of its presently unexplainable features, of
the type mentioned above. Such a theory inevitably would operate at
very short distances ($\sim M^{-1}_{\rm Planck}$) and very likely in
higher dimensions. It then becomes imperative, for reasons stated
above, that such a theory, as and when it evolves to be predictive,
should lead to an effective grand unification-like symmetry
(possessing SU(4)-color) in 4D near the string/GUT-scale, rather
than the SM symmetry.  If such a symmetry does emerge from the
underlying theory as {\em a preferred solution} in 4D, together with
the other desired properties mentioned above such as the presence of
three families with the (desired) hierarchical Yukawa couplings, it
would serve as {\em a very useful bridge} between the underlying
(string) theory and phenomenology.  It would explain observations in
the real world, beyond those encompassed by grand unification.
Needless to say, if such a solution would also explain the observed
cosmological constant (dark energy), one would be at a peak in the
path of understanding.

The picture depicted above is of course clearly {\em a wish and a
goal}, yet to be realized.  Entertaining such a wish amounts to
hoping that the current difficulties of string/M theory as regards
the large multiplicity of string vacua~\cite{ref:new} and lack of
selectivity (mentioned in the introduction) would eventually be
overcome possibly through a better understanding and/or formulation
of the theory, and most likely through the introduction of some
radically new ingredients,\footnote{Perhaps as radical as Bohr's
quantization rule that selected out a discrete set of orbits from an
unstable continuum, which in turn found its proper interpretation
within quantum mechanics.}

Entertaining such a hope no doubt runs counter to the recently
evolved view of landscape~\cite{ref:10a}, combined with
anthropism~\cite{ref:10b}. Such a hope is nevertheless inspired, as
mentioned in the introduction, by the striking successes we have had
over the last 400 years in our attempts at an understanding of
nature at a fundamental level.  To mention only a few that occurred
in the last 100 years, they include first and foremost the insights
provided by the two theories of relativity and quantum mechanics. In
the present context they include also the successes of the ideas of
the standard model, grand unification and inflation.  Each of these
have aided in varying degrees to our understanding of nature. {\em
At the same time, interestingly enough, each of these has provided
certain ingredients that are crucial to an understanding of the
evolution of life, in accord with anthropic constraints.}

Here, I have in mind, for example:  (i) The special theory of
relativity providing the  equation $E=mc^2$, which is crucial to an
understanding of energy-generation in stars; (ii) The radically new
laws of quantum mechanics, which are crucial to understanding the
stability of the atoms; (iii) The co-existence of quarks and
leptons, together especially with the charge-ratio $(Q_{e^-}
+Q_p)/Q_p \lsim 10^{-19}$, that are crucial to the formation of
atoms and equally important to the ``exact" neutrality of atoms and
thereby of the sun and the earth; (iv) The co-existence of the three
gauge forces---weak, electromagnetic and strong---each of which
plays a role in energy generation in the stars; note that both (iii)
and (iv) are neatly explained within grand unification, subject to a
testable assumption on symmetry breaking; (v) The existence of
neutrinos having  {\em (a) zero electric charge and (b)
non-vanishing but truly tiny masses $<$ 1 eV (with $m(\nu_3)/m_{\rm
top}\sim 10^{-12}$)}, as opposed to masses $\gg$ 50 eV (say); both
of these features are crucial to many stages of the cosmic drama
including  suitable structure formation and energy generation in
stars; and as we saw in Section 7 the tiny neutrino masses may well
be linked to the generation of baryon excess $Y_B \sim 10^{-10}$ in
the universe; all of these features involving neutrinos, which
certainly are in accord with  anthropic reasoning, are neatly
explained within the G(224) or SO(10)-unification (see Sections 4
and 7); (vi) The gross homogeneity and isotropy of the universe,
together with small density fluctuations, and simultaneously the
density parameter $\Omega \equiv \rho/\rho_c$ being so incredibly
close to the critical value of unity especially at early times ({\em
i.e.} $\Omega^{obs}_{\rm now} = \mathcal{O}(1)$ means $|
\Omega(t_{EW} \sim 10^{-11} s) - 1 | \lsim 10^{-27})$; both of these
features are crucial to understanding structure formation and origin
of life; they are both simply explained, however, by inflation and
quantum theory;  and last but not least, (vii) The existence of the
universal gravitational ``force", crucial to the evolution of the
universe leading to structure formation, which is dictated by string
theory.\footnote{The extreme weakness of the gravitational coupling
$G_N$ or correspondingly the large hierarchy between $M_{\rm
Planck}$ and $m_p$ remains, however, to be explained.}

These provide only a sample of examples; some of them involve very
small numbers like $(Q_{e^-}/Q_p+1) \lsim 10^{-19}$,
$m(\nu_3)/m_{\rm top} \sim 10^{-12}$, $Y_B \sim 10^{-10}$, and
$|\Omega(t_{EW})-1|\lsim 10^{-27}$. Each of these plays a crucial or
an important role in processes leading to the evolution of life as
we know it. Thus they are somehow ``needed" by (or are at least
compatible with) anthropic reasoning. At the same time, each of
these has a beautiful explanation in terms of an underlying
microscopic theory.\footnote{I should, however, mention that the
idea of inflation remains at present as a paradigm with strong
observational support.  Derivation of a suitable particle particle
theory model implementing this idea from an underlying theory like
string theory remains a challenge.} {\em These and many such
examples accumulating over 400 years suggest that anthropic
considerations, though clearly relevant, may well coexist with an
understanding of nature in the traditional sense, at least in most
cases, rather than being a substitute for it.}

Based on these features, {\em and} the fact that our current
understanding (including formulation) of string/M theory is rather
preliminary so as not to permit a real preference for the landscape
picture, I tend to lean towards the traditional approach to
understanding\footnote{Here I distinguish between understanding what
seems to be clearly microscopic phenomena (like the origin of the
three families and the smallness of the cosmological constant)
versus macroscopic ones such as the distance between the earth and
the sun.} \footnote{I should add that such a preference for
predictivity and traditional understanding of microscopic phenomena
can still be compatible with the multiverse picture and anthropism
as long as the set of vacuum solutions of the underlying theory (be
it either a dense discretum or a selective non-dense discrete set or
a sum of both, see text) contain solution(s) matching our own
world.} \footnote{On another note, regardless of how {\em our
sub-universe} was chosen ({\em e.g.} either statistically from a
large ensemble of a dense discretum or a very selective discrete
set), it is a remarkable fact that, so far, at least a very large
number of its features, though anthropically needed, have been
amenable to understanding in the traditional sense, as exemplified
above.  This has been possible only in terms of a few elegant
principles, some weird laws (like quantum mechanics), and a few
discrete choices. Even these (principles, laws and discrete choices)
which may vary from one sub-universe to another, may find a
rationale within an underlying theory.  {\em Since we cannot ever
tell if we have reached a dead-end in finding such principles and
weird laws and thereby a dead-end in the path of understanding,
unless and until the set of unexplainables (including the origin of
families and the cosmological constant) reduces to a null set, it
seems that striving towards reducing this set is a goal that is and
would remain worth pursuing, even within the multiverse picture.}}.
I believe that much can be gained by continuing to search for an
underlying {\em predictive} theory that includes quantum gravity.

It is of course possible that the underlying theory may in general
yield multiple solutions rather than a much desired unique or almost
unique one, for its ground state.  Even so, one needs to ascertain
(a) whether, in a more final picture,\footnote{In the context of
string/M theory, this should at least include  a non-perturbative
formulation of string/M-theory, including exploration of the
semi-perturbative inner region of the M-theory diagram.  As alluded
to in footnote s, the final picture (not yet in hand) may well and
most likely will involve radical departures from the current
framework.} these solutions constitute only a ``dense
discretum''~\cite{ref:new}---{\em i.e.} an almost continuum---or a
selective (non-dense) discrete set, or possibly a sum of  both, {\em
and most important, (b) whether they at least contain a sub-set of
solutions (however small) matching our own world, as regards its
known properties.}\footnote{This means having solutions which would
essentially yield the standard model at low energies (barring
entities such as neutrino masses as observed) {\em and} the desired
cosmological constant.} Until this latter feature emerges, it seems
to me that the underlying theory would remain questionable and the
landscape or statistically-based anthropic reasoning would not be
adequate anyway in accounting for our sub-universe.  {\em But if
such a feature does emerge, that in itself would be a monumental
achievement.}  Depending upon the nature of the solutions, for
instance if changing just one or a few parameters of
compactification would alter some or most of the observed parameters
so that the latter would not match the properties of our
sub-universe, then there can still be predictivity of some or most
of the observable parameters in terms of only a few. Testing these
predictions would test the underlying theory. This would greatly add
to our understanding.

This is the point of view amounting to a wish and a goal that I have
adopted in this talk as regards the origin of an effective grand
unification-like symmetry in 4D from an underlying theory.
Meanwhile, regardless of the origin of such an effective symmetry,
as noted in Sections 5, 6, and 8, it has ample predictions that are
amenable to experimental tests in the near future, should we succeed
in building the necessary facilities.

\section*{\large Acknowledgement}

I would like to thank Kaladi S. Babu, James D. Bjorken, Savas
Dimopoulos, Alon Faraggi, Shamit Kachru, Andrei Linde, Michael
Peskin, John Schwarz, Ashoke Sen, Stephen Shenker, Helen Quinn,
Marvin Weinstein, Frank Wilczek and Edward Witten for most helpful
comments and discussions. I would also like to thank the organizers
of the conference, especially Swapna Mohapatra and T. Padmanabhan,
for their kind hospitality.  I am grateful to Sharon Jensen for her
kind patience in typing this manuscript.

\end{document}